\documentclass[twocolumn,aps,prl,showpacs,superscriptaddress]{revtex4}
\usepackage{graphicx}
\begin{document}
\author{Yaroslav Tserkovnyak}
\affiliation{Lyman Laboratory of Physics, Harvard University,
Cambridge, Massachusetts 02138}
\author{Steven H. Simon}
\affiliation{Lucent Technologies Bell Labs, Murray Hill, NJ 07974}

\title{Monte Carlo Evaluation of Non-Abelian Statistics}

\begin{abstract}
We develop a general framework to (numerically) study adiabatic braiding of
quasiholes in fractional quantum Hall systems.
Specifically, we investigate the Moore-Read (MR)
state at $\nu=1/2$ filling factor, a known candidate for
non-Abelian statistics, which appears to actually occur in nature.
The non-Abelian statistics of MR quasiholes is
demonstrated explicitly for the first time, confirming the
results predicted by conformal field theories.
\end{abstract}

\date{\today}

\pacs{73.43.-f,05.30.-d}


\maketitle

The quantum statistics of a system of identical particles
describe the effect of adiabatic particle interchange
on the many-body wave function. All
fundamental particles belong to one of two classes: those that
have their wave function unaffected by particle interchange
(bosons) and those whose wave function gets a minus sign
under permutation (fermions). In two
dimensions, it is known that a number of exotic types of
statistics can exist for particle-like collective
excitations. For example, elementary excitations of the Laughlin
fractional quantum Hall (FQH) states exhibit \textquotedblleft
fractional\textquotedblright\ statistics: The phase of the wave
function is rotated by an odd fraction of $\pi$
when two Laughlin quasiparticles (or quasiholes) are interchanged
\cite{Halperin,Arovas}. Even more exotic statistics can exist when
a system with several excitations fixed at given
positions is degenerate \cite{MooreRead}.  In such a
case, adiabatic interchange (braiding) of excitations can
nontrivially rotate the wave function within the degenerate
space. In general, these braiding operations need not commute,
hence the statistics are termed
\textquotedblleft non-Abelian\textquotedblright.
Remarkably, the Moore-Read (MR)
state, a state which is commonly believed \cite{Read52}
to describe observed FQH plateaus
at $\nu=5/2$ and $7/2$ (which correspond respectively to half
filling of electrons or holes in the first excited Landau level),
is thought to have such non-Abelian
elementary excitations \cite{MooreRead}.
Other possible physical realizations of
non-Abelian statistics have also been proposed
\cite{OtherNonabelian}. States of this type have been
suggested to be attractive for quantum computation \cite{Kitaev}.

In Ref.~\cite{Arovas}, in order to establish the nature of
the statistics of the Laughlin quasiholes, a
Berry's phase calculation was performed that explicitly kept track
of the wave-function phase as one quasihole was transported
around the other. Although approximations were involved in this
calculation, it nonetheless established quite convincingly the
fractional nature of the statistics. Unfortunately, it has not
been possible to generalize this calculation to explicitly
investigate statistics of the MR quasiholes \cite{MooreRead}.
Although there has been much study of
the statistics of the MR quasiholes in the framework of
conformal field theories (CFT),
it would be desirable to perform a direct
calculation analogous to that of Ref.~\cite{Arovas}.
The purpose of this paper is to
provide such a calculation, albeit numerically.
Furthermore, the approach developed here is readily applicable
to other FQH systems which are not easily
accessible to analytic investigations.

The evolution operator of a many-body system described by a
Hamiltonian $H(\lambda)$ is in principle determined by the
Schr\"{o}dinger equation. In general, $H(\lambda)$ itself can change in time
through dependence on some varying parameter $\lambda(t)$.
In such a case, let us define $\varphi_i(t)$ at a given time $t$ to be an
orthonormal basis for a particular degenerate subspace,
requiring that this basis is locally smooth as a function of $t$. If
$\lambda$ is varied adiabatically
(and so long as the subspace does not cross any other states),
then the time-evolution operator maps
an orthonormal basis of the
subspace at one $t$ onto an orthonormal basis at
another $t$. A solution of the Schr\"{o}dinger equation,
$\psi_i(t)=U_{ij}(t)\varphi_j(t)$, is simply given by
\cite{WilczekZee}
\begin{equation}
(U^{-1}\dot{U})_{ij}=\,\,\left\langle\varphi_i\left|
\rule{0pt}{9pt}
\dot{\varphi}_j \right.\right\rangle\,\,\,\equiv A_{ij}(t)\,. \label{Ue}
\end{equation}
Since the matrix $A$ is anti-Hermitian, $U(t)$
is guaranteed to be unitary if its initial
value $U(0)$ is unitary.
Note that if we vary $\lambda$ so that the Hamiltonian returns to its
initial value at time $t$, i.e., $H(\lambda(t))=H(\lambda(0))$,
the corresponding transformation of the
degenerate subspace can be nontrivial, i.e.,
$\psi_i(t)\neq\psi_i(0)$ \cite{WilczekZee}.

We explicitly demonstrate that this is the case for the MR state with
at least four quasiholes.
The analysis is done in spherical geometry \cite{Haldane}:
$N$ electrons are positioned on a sphere of unit radius, with
their coordinates given by $(u_1,v_1), \ldots, (u_N,v_N)$,
using the spinor notation (i.e., $u=e^{i\phi/2}\cos\theta/2$ and
$v=e^{-i\phi/2}\sin\theta/2$ in terms of the
usual spherical coordinates). A monopole of charge $2S=2N+n-3$
in units of the flux quanta $\Phi_0=hc/e$ is placed in the
center of the sphere, giving rise to $2n$ quasiholes which are put at
$(\tilde u_1, \tilde v_1),\ldots,(\tilde u_{2n}, \tilde v_{2n})$.
Using gauge $\vec{A}=(\Phi_0S/2\pi)\hat{\phi}\cot\theta$, the
MR wave function \cite{MooreRead} is then given by
\begin{equation}
\psi_{\text{Pf}}=
\text{Pf}\Lambda^{(a,b,\ldots)(\alpha,\beta,\ldots)}_{ij}\prod_{i<j}(u_iv_j-v_iu_j)^2\,,
\label{Pf}
\end{equation}
where $\text{Pf}\Lambda^{(a,b,\ldots)(\alpha,\beta,\ldots)}_{ij}$
is the Pfaffian \cite{MooreRead} of the $N\times N$ antisymmetric matrix \cite{endnote1}
\begin{eqnarray} \nonumber
 \Lambda^{(a,b,\ldots)(\alpha,\beta,\ldots)}_{ij} =
(u_iv_j-v_iu_j)^{-1} &\times& \\
 \nonumber
\mbox{\Large[}(u_i\tilde{v}_a-v_i\tilde{u}_a)(u_j\tilde{v}_\alpha-v_j\tilde{u}_\alpha)
 &\times& \\ \nonumber
(u_i\tilde{v}_b-v_i\tilde{u}_b)(u_j\tilde{v}_\beta-v_j\tilde{u}_\beta)&\times&
\cdots \,\,\, + (i\leftrightarrow j) \mbox{\Large]}\,.
\end{eqnarray}
Pfaffian wave functions (\ref{Pf}) were first constructed in
Ref.~\cite{MooreRead} as CFT conformal blocks.
This MR state is the \textit{exact} ground state for a special
three-body Hamiltonian \cite{Greiter} and is also thought
to pertain for realistic two-body interactions in the first excited Landau
level \cite{Read52}.
The presence of quasiholes in the ground state is dictated by the
incommensuration of the flux with the electron number.
Physically, the MR state can be thought of as
p-wave BCS pairing of composite fermions (CF's) at zero net
field with quasiholes being the
vortex excitations \cite{MooreRead,ReadGreen,endnote2}.
Each quasihole has charge $e/4$ and corresponds to half a quantum of flux
(because of the paired order parameter \cite{MooreRead}).
Eq.~(\ref{Pf}) describes a state with quasiholes created in
two equal-size groups:
$(\tilde{u}_a,\tilde{v}_a),(\tilde{u}_b,\tilde{v}_b),\ldots$ and
$(\tilde{u}_\alpha,\tilde{v}_\alpha),(\tilde{u}_\beta,\tilde{v}_\beta),\ldots$.
Different quasihole groupings realize a space with degeneracy $2^{n-1}$
\cite{NayakWilczek,RezayiRead}. (Even though there are
$2n!/2(n!)^2$ ways to arrange $2n$ quasiholes into 2 groups of
$n$, the resulting wave functions are not all linearly independent.)
In the presence of finite-range interactions, the exact degeneracy may
be split by an amount exponentially small in the large vortex separation
\cite{ReadGreen}. In this case, infinitely
slow braiding will not exhibit non-Abelian statistics, although for a
very wide range of intermediate time scales, such statistics should
apply \cite{ReadGreen}. The effects of disorder on the statistics are only
partially understood \cite{ReadGreen}.

Consider an orthonormal basis $\varphi_i$, with
$i=1,\ldots,2^{n-1}$, for the subspace with $2n$ quasiholes, which
is locally smooth when parameterized by the quasihole coordinates.
In order to determine the braiding statistics, we find the transformation
$\varphi_i\rightarrow U_{ij}\varphi_j$ under the evolution
operator after two of the quasiholes are interchanged while the
others are held fixed. The unitary matrix $U_{ij}$
is obtained by first solving Eq.~(\ref{Ue}) and then projecting
the final basis onto the initial one. (Since we require
$\varphi_i$ to be only locally smooth, the basis itself can
nontrivially rotate after the quasiholes return to their
original positions). Eq.~(\ref{Ue}) is integrated numerically: The
differential equation is discretized
and the wave-function overlaps (the right-hand side of the
equation) are evaluated using the Metropolis Monte Carlo method.
The computational errors are
easily evaluated by varying the number of operations. We aim the
calculation at addressing the following questions: (1) What is the
Berry's phase accumulated upon quasihole interchange due to the
enclosed magnetic flux and due to the relative statistics?
(2) What is the transformation matrix for the ground-state subspace
corresponding to the braiding operations? In the following, we
will first describe the numerical method, then present the results,
and compare them to CFT predictions \cite{MooreRead,NayakWilczek}.

In order to integrate Eq.~(\ref{Ue}) numerically, the quasihole
interchange is performed in a finite number of steps. If $U^{(l)}$
is the value of the transformation matrix at the $l$th step, then
at the next step
\begin{equation}
U^{(l+1)}=U^{(l)}[1+A^{(l)}/2][1-A^{(l)}/2]^{-1}\,,
\label{Ul}
\end{equation}
where
$A^{(l)}_{ij}=\langle\varphi^{(l+1)}_i+\varphi^{(l)}_i|\varphi^{(l+1)}_j
-\varphi^{(l)}_j\rangle/2$. Our choice of the
finite-element scheme (\ref{Ul}) will become clear later. In
practice, in general we do not know an orthonormal basis for the
MR states (\ref{Pf}) in an analytic form, but we can numerically
orthonormalize a set of $2^{n-1}$ linearly-independent Pfaffian
wave functions $\psi_{\text{Pf}i}$. Let
$B^{(l)}_{ij}=[\psi^{(l)}_{\text{Pf}i},\psi^{(l)}_{\text{Pf}j}]$
denote the normalized overlaps of different states. (It is implied
here and throughout the paper that $[\psi^{(k)}_{\text{Pf}i},
\psi^{(l)}_{\text{Pf}j}]\equiv\langle\psi^{(k)}_{\text{Pf}i}|\psi^{(l)}_{\text{Pf}j}\rangle/\|\psi^{(k)}_{\text{Pf}i}\|\|\psi^{(l)}_{\text{Pf}j}\|$
is evaluated numerically.) We then easily show that
\begin{equation}
A^{(l)}=[V^{(l)}]^\dagger W^{(l)}V^{(l+1)}/2-\text{H.c.}\,,
\label{Al}
\end{equation}
where $W^{(l)}_{ij}=[\psi^{(l)}_{\text{Pf}i},\psi^{(l+1)}_{\text{Pf}j}]$
and $V^{(l)}$ is defined by $[V^{(l)}]^\dagger B^{(l)}V^{(l)}=\hat{1}$,
constructing an orthonormal basis
$\varphi^{(l)}_i=V^{(l)}_{ji}\psi^{(l)}_{\text{Pf}j}$.
We require $V^{(l)}$ to be locally smooth as a function
of the quasihole coordinates: The basis
 can continuously transform while the
quasiholes are moved, but, e.g., sudden sign flips are not
allowed.

According to Eq.~(\ref{Al}), $A^{(l)}$ is anti-Hermitian, so that
the transformation $U^{(l+1)}$ is guaranteed to be
unitary if $U^{(l)}$ is unitary. This explains our choice
(\ref{Ul}) for discretizing Eq.~(\ref{Ue}). Another feature preserved
by our numerical scheme is that making a step forward,
$\psi^{(l)}_{\text{Pf}i}\rightarrow\psi^{(l+1)}_{\text{Pf}i}$,
followed by a step backward,
$\psi^{(l+1)}_{\text{Pf}i}\rightarrow\psi^{(l)}_{\text{Pf}i}$,
results in a trivial transformation.
We start at $U^{(0)}=\hat{1}$ and find $U^{(n_s)}$ after performing
$n_s+1$ steps for braiding of two quasiholes ($n_s$ is increased to
convergence). Because $\psi^{(n_s)}_{\text{Pf}i}$ is some
nontrivial linear combination of $\psi^{(0)}_{\text{Pf}i}$, we,
finally, have to project the transformation onto the initial
basis: $U^{(n_s)}\rightarrow U^{(n_s)}O^{T}$, where
$O=[V^{(0)}]^\dagger\Omega V^{(n_s)}$ and
$\Omega_{ij}=[\psi^{(0)}_{\text{Pf}i},\psi^{(n_s)}_{\text{Pf}j}]$.
The resulting unitary transformation matrix $U$ then gives a
representation of the braid group for quasihole interchanges. In
the following, we describe our numerical experiments.
\begin{figure}
\includegraphics[width=8.1cm,clip=]{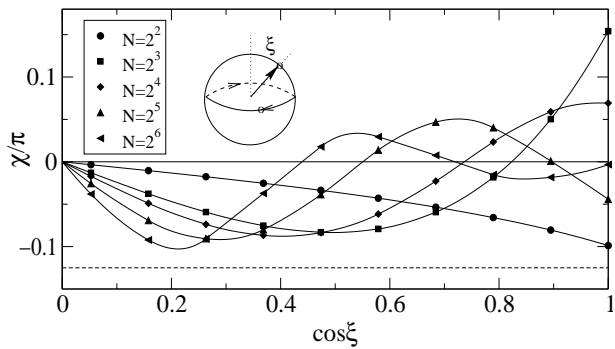}
\caption{\label{fig1} Berry's phase $\chi$ for looping one MR
quasihole around the equator with another quasihole fixed at a
zenith angle $\xi$. $N=4,8,16,32,64$ is the number of electrons.
The dashed line, $\chi/\pi=-1/8$,
shows a naive prediction. For $\cos\xi\approx 0$,
the two quasiholes approach each other very closely and we see strong
finite-size oscillations in the Berry's phase. For larger $N$ and
$\cos\xi$ (i.e., larger quasihole separation in units of the magnetic
length), $\chi$ appears to be converging toward zero.
$\chi(-\cos\xi)=-\chi(\cos\xi)$.}
\end{figure}

The space describing $2n=2$ MR quasiholes is nondegenerate, so
non-Abelian statistics cannot occur.  There is, nevertheless, a
Berry's phase accumulated from wrapping these quasiholes around
each other. Our calculation of this phase for the MR state is
analogous to the one performed in Ref.~\cite{Arovas} for the
Laughlin state, except that our calculation is numerical and
therefore requires no mean-field approximation. Let us first
briefly recall results for the Laughlin wave function at filling
factor $\nu=1/p$.
In the disk geometry, the Berry's phase $\chi$ corresponding to
taking a single quasihole around a loop is given by $2\pi$ for
each enclosed electron, i.e., $\chi=2\pi\langle N\rangle$, where
$\langle N\rangle$ is the expectation number of enclosed
electrons \cite{Arovas}.
Therefore, when another quasihole is moved inside the loop, the
phase $\chi$ drops by $2\pi/p$ which implies fractional statistics
of the quasiholes. In spherical
geometry \cite{Haldane}, the same result holds unless the south
and north poles (which have singularities in our choice of
gauge) are located on different sides of the loop.
In the latter case, the Berry's phase is given by $\chi=\pi\langle
N_{\text{in}}-N_{\text{out}}\rangle$, where
$N_{\text{in}(\text{out})}$ is the number of electron inside
(outside) the loop. If a single Laughlin quasihole is
then looped around the equator, its Berry's phase vanishes, but if
another quasihole is placed above or below, the phase becomes
$\chi= \pm \pi/p$.  We check our Monte Carlo method by reproducing
these results numerically. The charge of the MR
($\nu=1/2$) quasihole is $e/4$, so that by analogy with the
Laughlin state one might naively expect that the Berry's phase for
looping one quasihole around the equator with another fixed above
or below it is given by $\chi=\pm\pi/8$ \cite{Greiter} (with an extra factor
of $1/2$ due to MR quasiholes corresponding to only half of the flux quantum).
In Fig.~\ref{fig1} we show numerical calculation of
$\chi$ for a MR system having 2 quasiholes,
one looped around the equator and the other
held fixed. If the two quasiholes approach each other too
closely, we see strong finite-size oscillations in the Berry's
phase. However, for larger separation, $\chi$
appears to be converging towards zero, which
was first predicted in Ref.~\cite{ReadMoore}
and can be well understood using the plasma analogy \cite{Gurarie}.
\begin{figure}
\includegraphics[width=8.1cm,clip=]{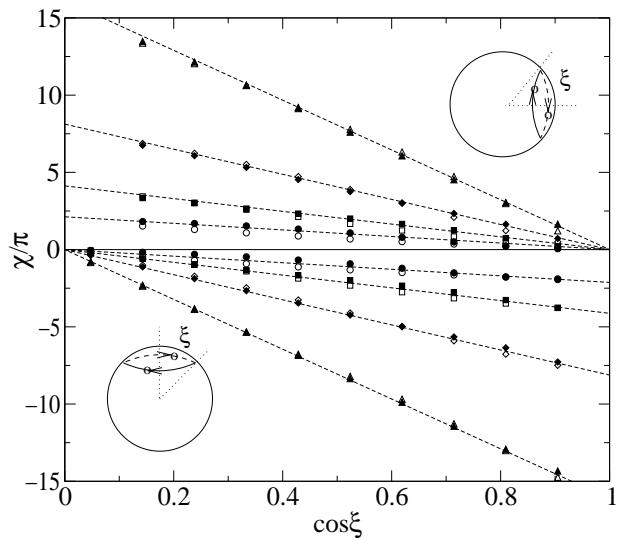}
\caption{\label{fig2} For $\chi>0$ ($\chi<0$) filled symbols
show the phase accumulated by interchanging two quasiholes around
a circle with opening angle $\xi$ centered on the equator
(north pole), for various $N$ as in Fig.~\ref{fig1}.
The straight dashed lines in the upper half are
$0.5(N+1/4)(1-\cos\xi)$, corresponding to the expectation of the
number of electrons enclosed by the loop.
The $+1/4$ accounts for the charge pushed out by one of the quasiholes.
For $\chi<0$, the dashed lines are
$-0.5(N+1/4)\cos\xi$, i.e., one half of the number of
electrons inside minus one half the number outside the loop. Open
symbols, corresponding to a similar calculation with one
quasihole moving and the other fixed at the center of the circle,
almost overlay the filled symbols, confirming the trivial relative statistics.}
\end{figure}
\begin{figure}
\includegraphics[angle=-90,width=8.1cm,clip=]{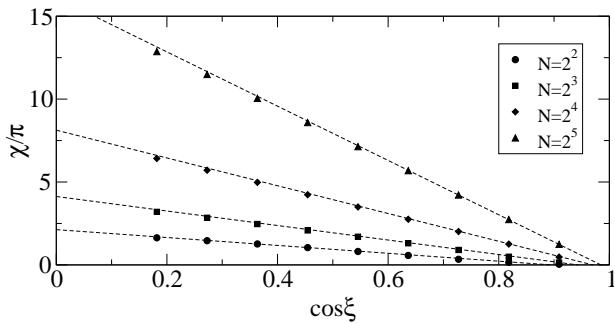}
\caption{\label{fig3}Same as the upper half of Fig.~\ref{fig2},
but now with four quasiholes present, two of which are fixed on
the equator, at $\phi=\pm3\pi/4$, and two interchanged, with
initial and final positions at $\phi=\pm\xi$ on the equator. The
straight dashed lines are $0.5(N+3/4)\cos\xi-1/4$. Here, $+3/4$
accounts for the average electron-density correction for the
charge localized at $2n-1$ quasiholes. The additional phase offset
of $-1/4$ reflects the Abelian part of the braiding statistics, in
agreement with the predictions of Refs.~\cite{MooreRead,NayakWilczek}.}
\end{figure}

Even though the relative statistics of two MR quasiholes are
trivial, they do pick up a phase due to their wrapping around the
electrons, analogous to what occurs in the Laughlin case.
Fig.~\ref{fig2} shows that as the size of the system increases,
the phase accumulated by interchanging two quasiholes (filled symbols)
or braiding one around the other (open symbols)
can be well approximated by assuming the wave
function rotates by $\pi$ for each enclosed electron (compare to
$2\pi$ for the Laughlin state), when the poles are not separated
by the loop (and the effect of the pole singularities is
analogous to that in the Laughlin state). Even for
systems consisting of only 4 electrons, this approximation stays
quite good if we correct the average electron density for the
charge pushed out by one localized quasihole (see dashed lines
in Fig.~\ref{fig2}).  This method of correcting the average
density also works for the Laughlin state on the
sphere.

\begin{figure}[b]
\includegraphics[angle=-90,width=8.1cm,clip=]{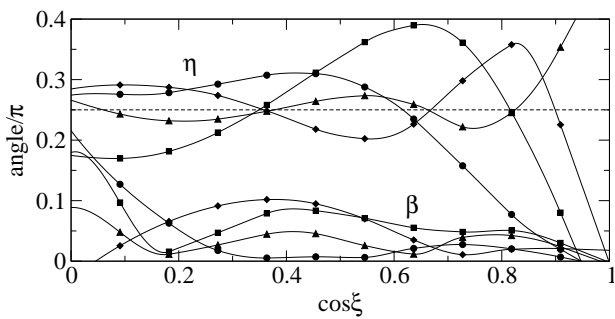}
\caption{\label{fig4}Parameters $\eta$ and $\beta$ defining
transformation matrix (\ref{U}) for the same operations as $\chi$
shown in Fig.~\ref{fig3}. The dashed line shows $1/4$, an
approximation used for $\eta$ in the text. Similarly $\beta$ can be
approximated as zero [so that $\epsilon$ in Eq.~(\ref{U}) is not
defined].
These approximations become better with larger system
size and for intermediate $\cos \xi$ when the quasiholes remain
further apart.
The symbol convention is the same as in
Fig.~\ref{fig3}.  Lines interpolate Monte Carlo results.}
\end{figure}

We now turn to $2n=4$ MR quasiholes, which is the
simplest case when statistics can be non-Abelian (the
ground state has degeneracy 2).
While the above results for 2 quasiholes are anticipated by the plasma
analogy \cite{Gurarie}, one may need deeper CFT
\cite{MooreRead,NayakWilczek} arguments
in order to understand the following findings.
In the calculation,
we first fix all quasiholes on the equator and then interchange an
adjacent pair of them around a circle with different opening
angles $\xi$ centered on the equator. Parameterizing a
unitary matrix $U$ by
\begin{equation}
U=e^{i\chi}\left(
\begin{array}{cc}
e^{i\eta}\cos\beta/2 & ie^{-i\epsilon/2}\sin\beta/2 \\
ie^{i\epsilon/2}\sin\beta/2 & e^{-i\eta}\cos\beta/2
\end{array}
\right)\,,
\label{U}
\end{equation}
we plot in Figs.~\ref{fig3} and \ref{fig4} the results
(in a convenient basis) for the transformation
$U_1$ corresponding to the braiding operation on one of the
quasihole pair. Due to the rotational symmetry around the
vertical axis, knowing $U_1$ we can deduce other
transformations $U_2$, $U_3$, and $U_4$ (for interchanges of
pairs ordered along the equator) by rotating and
projecting the initial basis and correspondingly transforming
$U_1$. It is then easy to show that $U_1=U_3$ and $U_2=U_4$ due to
the form (\ref{Pf}) of the wave function. Furthermore, we find
numerically that $U_2\approx F^\dagger U_1F$, where $
F=(\sigma_z-\sigma_x)/\sqrt{2}$, $\sigma$'s being the
usual Pauli matrices. This approximation is good within a few
percent for smaller systems and is even better for larger
ones.

According to Fig.~\ref{fig4}, we see that apart from the Abelian
phase $\chi$, $U_1$ can be approximated by
$U_1\approx\text{diag}(1+i,1-i)/\sqrt{2}$, with the disagreement
becoming smaller for larger systems. Using $F$, we can then
construct all other matrices $U_i$. After
performing the above approximations, we find that the unitary
transformations corresponding to the braid operators realize the
right-handed spinor representation of SO$(2n)\times$U$(1)$
(restricted to $\pi/2$ rotations around the axes) as
predicted in Ref.~\cite{NayakWilczek} using CFT.  In addition to
the usual relations required of a representation of the braid
group on the plane, on the sphere the generators must obey an
additional relation. For the case of $2n=4$, for
example, we expect to have $U_1 U_2 U_3 U_3 U_2 U_1=1$.  One can
easily show that (for general $n$) the relevant representation of
the braid group predicted in Ref.~\cite{NayakWilczek} satisfies
this additional relationship up to an Abelian phase. (The
failure of the Abelian phase to satisfy this law is related to the
gauge singularities, and will be discussed elsewhere.)

In summary, we formulated a numerical method to study
braiding statistics of FQH excitations and applied it to
perform the first direct calculation of the non-Abelian statistics
in the MR state. Our findings confirm results previously
drawn within the CFT framework.

We have enjoyed helpful discussions with B. I. Halperin, C. Nayak, F. von
Oppen, and N. Read. This work was supported in part by NSF Grant DMR 99-81283.
The computations were performed on the O2000 supercomputer CPU farm
at Boston University.


\begin{thebibliography}{20}
\bibitem{endnote1} Computationally, the evaluation of the Pfaffian is not very
expensive since $(\text{Pf}\Lambda)^2=\text{Det}\Lambda$, the
matrix determinant. The sign of the square root can be obtained by
enforcing appropriate linear relations for the overcomplete basis
of Pfaffian wave functions \cite{NayakWilczek}.
\end{thebibliography}
\end{document}